\documentclass[journal,twoside,web]{ieeecolor}
\usepackage{lcsys}
\usepackage{cite}
\usepackage{amsmath,amssymb,amsfonts}
\usepackage{bbold}
\usepackage{mathtools}
\usepackage{algorithmic}
\usepackage{graphicx}
\usepackage{textcomp}
\usepackage{hyperref}
\def\BibTeX{{\rm B\kern-.05em{\sc i\kern-.025em b}\kern-.08em
    T\kern-.1667em\lower.7ex\hbox{E}\kern-.125emX}}
\begin{document}
\title{Deep Reinforcement Learning for Market Making Under a Hawkes Process-Based Limit Order Book Model}

\author{Bruno Gašperov and Zvonko Kostanjčar, \IEEEmembership{Member, IEEE}
\thanks{This work was supported in part by the Croatian Science Foundation under Project 5241, and in part by the European Regional Development Fund under Grant KK.01.1.1.01.0009 (DATACROSS).}
\thanks{The authors are with the Laboratory for Financial and Risk Analytics, Faculty of Electrical Engineering and Computing, University of Zagreb, 10000 Zagreb, Croatia (e-mail: bruno.gasperov@fer.hr; zvonko.kostanjcar@fer.hr).}
}
\onecolumn
\noindent  978-1-5386-5541-2/18/\$31.00~\copyright2018 IEEE \\
\\
This paper is a preprint version of the paper: \\
Gašperov, Bruno, and Zvonko Kostanjčar. "Deep Reinforcement Learning for Market Making Under a Hawkes Process-Based Limit Order Book Model." \emph{IEEE Control Systems Letters} 6 (2022): 2485-2490, DOI: 10.1109/LCSYS.2022.3166446, \url{https://ieeexplore.ieee.org/document/9754690}
and it is IEEE copyrighted material. For IEEE copyright policy please refer to \url{https://www.ieee.org/publications/rights/copyright-policy.html}
\twocolumn
\maketitle
\thispagestyle{empty}
\begin{abstract}
The stochastic control problem of optimal market making is among the central problems in quantitative finance. In this paper, a deep reinforcement learning-based controller is trained on a weakly consistent, multivariate Hawkes process-based limit order book simulator to obtain market making controls. The proposed approach leverages the advantages of Monte Carlo backtesting and contributes to the line of research on market making under weakly consistent limit order book models. The ensuing deep reinforcement learning controller is compared to multiple market making benchmarks, with the results indicating its superior performance with respect to various risk-reward metrics, even under significant transaction costs.
\end{abstract}


\section{Introduction}
\label{sec:introduction}
\IEEEPARstart{O}{ptimal} market making (MM) is the problem of placing bid (buy) and ask (sell) orders simultaneously on both sides of the limit order book (LOB) with the goal of maximizing trader's terminal wealth, all while minimizing the associated risks. Arguably the most salient among them is the inventory risk, which stems from price movements in the underlying asset held by the trader (market maker) in its inventory. A risk-averse market maker typically prefers to keep its inventory consistently close to zero. In order to achieve this, continuous adjustment of prices at which the orders are placed (controls) in response to the current inventory and other relevant variables, is required. This gives rise to a natural formulation of MM as a stochastic optimal control problem. Within the most commonly used framework, first used in the seminal Avellaneda-Stoikov (AS) paper \cite{as} and later studied and generalized by a plethora of researchers \cite{fodra}--\cite{cartea}, the problem boils down to reducing the associated Hamilton-Jacobi-Bellman equation to a system of ordinary differential equations (ODE) and finally deriving closed-form approximations to the optimal MM controls. Alternatively, near-optimal controls could be obtained by techniques such as (deep) reinforcement learning.

Reinforcement learning (RL) is a class of trial-and-error-based methods for sequential decision making under uncertainty, used to solve stochastic optimal control problems framed as Markov decision processes (MDP). RL methods are particularly powerful when combined with deep neural networks (which are used as function approximators), coalescing into a novel field of deep reinforcement learning (DRL). Recent years have witnessed a multitude of striking DRL achievements in a diverse set of domains ranging from the game of Go \cite{silver}, to text generation \cite{ranzato} and networked control systems \cite{burak}. Unsurprisingly, there is nowadays a surge of interest in the applications of (D)RL to the problem of MM, which is naturally framed as a discrete-time MDP. 

In \cite{lim}, the authors present what they claim to be the first practical application of RL to optimal MM in high-frequency trading, using the LOB model proposed in \cite{cont} to train an RL agent via simple discrete Q-Learning. The authors then demonstrate that their framework outperforms the AS and fixed-offset benchmarks. Somewhat similarly, Spooner et al. \cite{spooner} use the SARSA algorithm with a linear combination of tile codings as a value function approximator to produce an RL agent demonstrating superior out-of-sample performance across multiple securities. Sadighian \cite{sadighian} develops a framework for cryptocurrency MM employing two advanced policy gradient-based RL algorithms (A2C and PPO) and a state space comprised of both LOB data and order flow arrival statistics. Gašperov and Kostanjčar \cite{gasperov} present a framework underpinned by ideas from
adversarial RL and neuroevolution, with experimental results demonstrating its superior reward-to-risk performance. Other notable approaches introduce additional features like the presence of a dark pool \cite{baldacci}, multi-asset MM in corporate bonds \cite{gueant2}, and dealer markets \cite{ganesh}. 

Regardless of the approach used, faithful LOB modeling, ideally accounting for the empirical properties and stylized facts of market microstructure as well as the discrete nature of the LOB itself, is pivotal to obtaining high-performing MM controllers. However, due to the naive assumptions they are predicated upon, the LOB models underlying most contemporary MM approaches remain inconsistent with respect to direction, timing, and volume, leading to phantom gains under backtesting and preposterous events \cite{law}, such as price decreases after a large buy market order. For example, in the original AS model \cite{as}, price movements are assumed to be completely independent of the arrivals of market orders and the LOB dynamics, while the subsequent approaches only partly address such inconsistencies. To ameliorate this, a novel weakly-consistent pure-jump market model that ensures that the price dynamics are consistent with the LOB dynamics with respect to direction and timing is proposed in \cite{law}. Nevertheless, it still assumes constant order arrival intensities, meaning that any (empirically found) effects of self- or mutual- excitation and inhibition between various types of LOB order arrivals remain unaccounted for.

In this paper, we consider the stochastic optimal control problem of a market maker trading in a weakly consistent, multivariate Hawkes process-based LOB model. To the authors' knowledge, our approach is the first to wed DRL to MM under such a model. On one hand, our goal is to increase the realism of existing MM models, such as the model proposed in \cite{law}. Given that Hawkes processes are commonly used for realistic modeling of market microstructure \cite{bacry, luabergel1, fonseca}, we opt for such a model, which is then used as a backbone of our simulator. On the other hand, we also need to ensure not to jeopardize tractability. To this end, we avoid the use of full LOB models (because of the associated complexities and issues \cite{law}) and choose a reduced-form LOB model instead. Furthermore, a number of additional simplifying assumptions about microstructural dynamics are made. Therefore, underpinning our work is a careful consideration of the balance between complexity and tractability. The model is also interactive in the sense that the market maker's controls (choice of placed limit and market orders) affect the order arrival intensities, i.e., the market responds dynamically to the market maker's control strategy. The proposed approach contributes to the (in our view) underrepresented line of research on MM under weakly consistent LOB models \cite{law}. It also enables the leveraging of the advantages of Monte Carlo backtesting, particularly its ability to perform controlled randomized experiments \cite{deprado}. 

\section{Preliminaries}

\subsection{Multivariate Hawkes processes}

A $p$-dimensional linear Hawkes process \cite{embrechts} is a $p$-dimensional point process $N(t)= (N_{k}(t):k= 1, \ldots, p )$ with the intensity of $N_k$ (the $k$-th dimension) given by:

\begin{equation}
\lambda_{k}(t)=\mu_{k}+\sum_{l=1}^{p} \int_{0}^{t-} f_{k,l}(t-s) \mathrm{d} N_{l}(s),
\end{equation}
where $\mu_{k} \geq 0$ are the baseline intensities, $N_l(t)$ the number of arrivals within $[0,t]$ corresponding to the $l$-th dimension, and $f_{k,l} (t)$ the kernels (triggering functions). Arrivals in dimension $l$ perturb the intensity of the arrivals in dimension $k$ at time $t$ by $f_{k,l}(t-s)$ for $t>s$. Multivariate linear Hawkes processes are relatively easy to handle and lend themselves well to simulation and interpretation due to their branching structure representation \cite{law2}. In cases when exponential kernels are used, the intensities are given by:

\begin{equation}
\lambda_{k}(t)=\mu_{k}+\sum_{l=1}^{p} \int_{0}^{t-} \alpha_{k,l} e^{-\beta_{k,l}(t-s)} \mathrm{d}N_{l}(s),
\end{equation}
where $\beta_{k,l} \geq 0$ and $\alpha_{k,l} \geq 0$ are the decay and excitation parameters.

\section{MODEL}

\begin{figure}[!t]
\centerline{\includegraphics[width=\columnwidth]{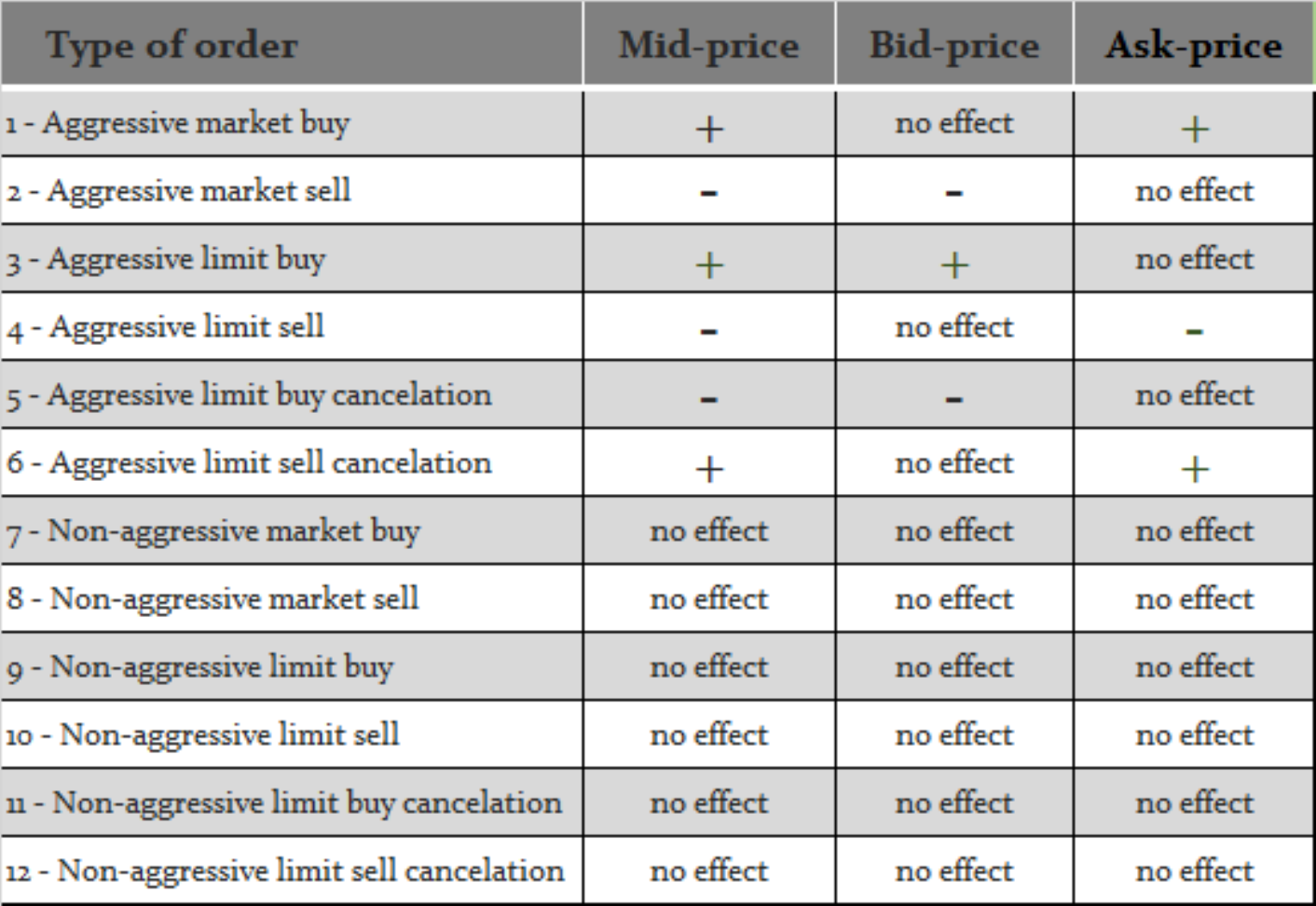}}
\caption{Types of orders and their effect on the mid-price, bid, and ask price.}
\label{fig:table1}
\vspace{-5ex}
\end{figure}

\subsection{Event types}
We employ the categorization of LOB events proposed by Biais \cite{biais}, comprising 12 different types of LOB events, as provided in Fig.~\ref{fig:table1}. Untypical orders, such as iceberg orders or (partly) hidden orders, are omitted from consideration for simplicity's sake. Aggressive market orders, limit orders, and cancellations (events of type 1--6) affect either the bid or the ask price, and consequently the mid-price as well. Non-aggressive market orders (type 7--8) do not affect any of the prices; however, they generate trades and affect the volume at the top of the LOB and are hence relevant to MM. Non-aggressive limit orders and cancellations (type 9--12) affect neither the prices nor the volume at the top of the LOB and are hence ignored. Consequently, we consider 8 event types in total: 
\begin{equation}
    E_{\text{all}} = \{ M_{b}^{a}, M_{s}^{a}, L_{b}^{a}, L_{s}^{a}, C_{b}^{a}, C_{s}^{a}, M_{b}^{n}, M_{s}^{n} \},    
\end{equation}
where the superscript denotes the (non)-aggressiveness, and the subscript the side (buy/sell). The resulting LOB model is weakly consistent following \cite{law}. Allowing for variable jump sizes, the LOB dynamics are modeled by an 8-variate marked point process, where marks indicate jump sizes. The corresponding counting process is given by:

\begin{equation}
N(t)=\left(N_{M_{b}^{a}}(t), \ldots, N_{M_{s}^{n}}(t)\right),
\end{equation}
and the associated intensity vector by:
\begin{equation}
\lambda(t)=\left(\lambda_{M_{b}^{a}}(t), \ldots, \lambda_{M_{s}^{n}}(t)\right).
\end{equation}
After denoting the jump size (in ticks) associated with an individual event $e$ by $J_e$, the mid-price $P_t$ is given by:
\begin{equation}
    P_t=P_0 + \left(\sum_{\substack{e,\ T(e) \in E_{\text{inc}}}}J_{e} \ - \sum_{\substack{e,\ T(e) \in E_{\text{dec}}}}J_{e} \right) \frac{\delta}{2},
\end{equation}
where $P_0$ is the initial price, $\delta$ the tick size, $T(e)$ the type of event $e$, $E_{inc}= \{ M_{b}^{a}, L_{b}^{a}, C_{s}^{a} \}$, and $E_{dec}= \{ M_{s}^{a}, L_{s}^{a}, C_{b}^{a} \}$. Jump sizes are, for the sake of simplicity, assumed to be independent of the jump times and i.i.d. (similarly as in \cite{law}).

\subsection{Simulation procedure}

A multivariate linear Hawkes process with exponential kernels is used to model dependencies, namely cross-excitation and self-excitation effects, between different LOB order arrivals (i.e.\ event types). The choice of exponential kernels is motivated by both their tractability to simulation and suitability for modeling short-term memory processes describing market microstructure, due to which they are used traditionally in financial applications \cite{sornette}\cite{simon}. Furthermore, their use conveniently results in the Markovian properties of the process \cite{bacry} \cite{oakes}. In order to simulate the multivariate Hawkes process, we rely on Ogata's modified thinning algorithm \cite{ogata}. 
\vspace{-2ex}
\subsection{Market making procedure}
The MM procedure roughly follows the one described in \cite{as}. At the start of each time-step, at time $t$, the agent (controller) cancels its outstanding limit orders (if there are any), and observes the state of the environment $S_t$ containing market and agent-based features. The agent uses this information to select the action $A_t$ - it decides whether to post limit orders (and at which prices) or a market order. If the absolute value of the agent's inventory is equal to the inventory constraint $c$, $c \in \mathcal{N}$, the order on the corresponding side is ignored. All relevant market (mid-price, bid and ask price, spread) and agent-based (inventory, cash) variables are then updated accordingly. Next, the LOB events generated by the simulation procedure are processed sequentially, which is followed by corresponding updates of the relevant variables. Executed limit orders cannot be replaced with new ones until the beginning of the next time-step. Finally, the agent reaches the end of the time-step and receives the reward $R_{t+\Delta t}$. When time $t+\Delta t$ is reached, unexecuted limit orders from the previous time-step are canceled, the agent observes the new state of the environment $S_{t+\Delta t}$, selects the action $A_{t+\Delta t}$ and the procedure is iterated until the terminal time $T$. The agent's inventory $I_t$ is described by the following relation:
\begin{equation}
    \mathrm{d}I_t = \mathrm{d}N_t^{\text{b}}-\mathrm{d}N_t^{\text{a}} + \mathrm{d}N_t^{\text{mb}} - \mathrm{d}N_t^{\text{ms}},
\end{equation}
where $N_t^{\text{b}}$, $N_t^{\text{a}}$, $N_t^{\text{mb}}$, and $N_t^{\text{ms}}$ denote the number of the agent's limit bid (buy), limit ask (sell), market buy, and market sell orders executed up to time $t$, respectively. The process $N_t^{\text{b}}$ is described by:
\begin{equation}
\mathrm{d}N_t^{\text{b}}= \mathrm{d}N_{M_{s}^{a}} \mathbb{1}_{\text{fill, } M_{s}^{a}} +\mathrm{d}N_{M_{s}^{n}}\mathbb{1}_{\text{fill, } M_{s}^{n}},
\end{equation}
where $\mathbb{1}_{\text{fill, } M_{s}^{a}}$ ($\mathbb{1}_{\text{fill, } M_{s}^{n}}$) is the indicator function for whether the incoming (non)-aggressive market order fills the market maker's limit order. The process $N_t^{\text{a}}$ is described analogously. Finally, the agent's cash process $X_t$ is given by:
\begin{equation}
    \mathrm{d}X_t = Q^{a}_t \mathrm{d}N_t^{\text{a}} - Q^{b}_t \mathrm{d}N_t^{\text{b}} - (P_{t}^{\text{a}}+\epsilon_t)\mathrm{d}N_t^{\text{mb}} + (P_{t}^{\text{b}}-\epsilon_t)\mathrm{d}N_t^{\text{ms}},
\end{equation}
where $Q^{a}_t$ ($Q^{b}_t$) denotes the price at which the agent's ask (bid) quote is posted, $P_{t}^{\text{a}}$ ($P_{t}^{\text{b}}$) the best ask (bid) price, and $\epsilon_t$ the additional costs due to fees and market impact, all at time $t$. Furthermore, a number of simplifying assumptions are made:
\begin{itemize}
    \item Market orders submitted by the market maker are aggressive with probability $Z_1$, and its limit order cancellations are aggressive with probability $Z_2$.
    \item Exponential distribution with density given by
    $f(x)=\frac{1}{\beta} \exp \left(-\frac{x-\mu}{\beta}\right)$ is used for modeling the size of the price jumps $J_e$ associated with aggressive events, where $\mu$ is the location and $\beta$ the scale parameter. Where required (i.e. with aggressive limit orders and cancellations), the truncated exponential distribution with the corresponding parameters is used.
    \item Upon arrival of a non-aggressive market order, the probability of execution of the market maker's limit order standing at the best bid/ask price is fixed and given by $Z_3$. The limit order is either executed in its entirety or not at all.
\end{itemize}

\section{REINFORCEMENT LEARNING CONTROLLER}

\subsection{State space}

The state space consists of the current inventory $I_t$, the current bid-ask spread $\Delta_t$, and the trend variable $\alpha_t$:
\begin{equation}
    S_t = (I_t, \Delta_t, \alpha_t).
\end{equation}
Due to the inventory constraints, there are $2c+1$ possible inventory states, i.e., $I_t \in \{-c, \dots, c\}$. 
The current bid-ask spread $\Delta_{t}$ is measured in ticks and strictly positive. 
The variable $\alpha_t$ accounts for the trend and is calculated as $\alpha_t = \lambda_{M_{b}^{a}}(t)+\lambda_{M_{b}^{n}}(t)-\lambda_{M_{s}^{a}}(t)-\lambda_{M_{s}^{n}}(t)$. The inventory feature $I_t$ is normalized by the min-max normalization. Since features $\Delta_t$ and $\alpha_t$ have unknown means and variances, we generate a long trajectory with $100{\small,}000$ steps by a controller that outputs completely random actions and use the obtained means and variances for $z$-score normalization. 
\vspace{-2ex}
\subsection{Action space}

The action (i.e.\ the controls) $A_t$ at time $t$ corresponds to a pair of offsets from the best bid (ask) prices. Hence:

\begin{equation}
A_t = (Q^{a}_t-P_{t}^{\text{a}}, P_{t}^{\text{b}} - Q^{b}_t).
\end{equation} 
The agent is allowed to post limit orders at all possible prices, thereby determining the level of aggressiveness/conservativeness on each of the sides of the LOB. If the agent posts a limit order pair with a negative quoted spread, the action is ignored. A bid (ask) order posted at or above (below) the best ask (bid) is treated as a buy (sell) market order and executed immediately. 
All limit and market orders sent by the market maker are assumed to be of unit size (i.e. are for one unit of the asset) and the controls are rounded to a multiple of the tick size. 
\vspace{-2.18ex}
\subsection{Reward function}

The goal of the MM agent is to maximize the expectation of the terminal wealth while simultaneously minimizing the inventory risk. We take inspiration from the commonly used MM formulation described in \cite{cartea} and assume that the market maker maximizes the following expression:

\begin{equation}
\mathrm{E}_{\pi}\left[W_{T}-\phi \int_{0}^{T} |I_{t}| d t \right], 
\end{equation}
over the set of RL policies. Each policy $\pi : \mathcal{S}  \to \mathcal{P}(\mathcal{A})$ maps states to probability distributions over the action space. $W_t = I_t P_t + X_t$ is the total wealth at time $t$, $T$ the terminal time, and $\phi \geq 0$ the running inventory penalty parameter which is used to disincentivize the market maker from holding non-zero positions and thereby exposing itself to the inventory risk. Note that, unlike in \cite{cartea}, we use the absolute value of the inventory instead of the quadratic inventory penalty to obtain a convenient value at risk (VaR) interpretation. 
Therefore, the reward at time $t+\Delta t$ is given by:
\begin{equation}
R_{t+\Delta t} = \Delta W_{t+\Delta t} - \phi \int_{t}^{t+\Delta t} |I_{s}| d s.
\end{equation}
The integrand is piecewise constant and hence trivial. 
\vspace{-2ex}
\subsection{Controller design}
A neural network with 2 fully-connected hidden layers of 64 neurons with ReLU activation is used to represent the controller. Such a relatively shallow architecture is somewhat of a standard in DRL and, moreover, simple NN architectures have been shown before to perform comparably to (or better than) sophisticated CNN-LSTM architectures for modeling of the LOB \cite{briola}.
\vspace{-2ex}
\subsection{Training}
SAC (Soft Actor-Critic) is used to train the RL controller in our experiments. SAC is a state-of-the-art actor-critic RL algorithm capable of tackling continuous action spaces, and characterized by increased robustness and the ability to learn multiple modes of near-optimal behavior. It is a maximum entropy algorithm, meaning that it maximizes not only the return but also the policy entropy, thereby improving exploration. Alternatively, DQN might be used; however, our experimentation showed that it suffers from severe instability issues, especially in its vanilla variant. TD3 (Twin Delayed DDPG) has also been considered but was eventually excluded due to its inferior performance. The number of training time-steps is set to $10^{6}$.
\vspace{-2ex}
\section{EXPERIMENTS}

In order to pit the performance of our DRL approach against MM benchmarks, we perform Monte Carlo simulations (backtests) using synthetic data generated by the simulator. The number of simulations is set to $10^3$. Standard MM benchmarks like the AS \cite{as} approximations are ill-suited for our framework since they take into account neither the existence of the bid-ask spread nor the discrete nature of the underlying LOB \cite{gueantbook}. We hence need to turn to alternative benchmarks. We consider a class of MM strategies linear in inventory and including inventory constraints. Among the members of this class we denote the best performing one as the LIN strategy. Note that this class of strategies also includes the state-of-the-art Gueant-Lehalle-Fernandez-Tapia (GLFT) \cite{gueant} approximations. Additionally, we consider a simple strategy (SYM strategy) that always places limit orders precisely at the best bid and the best ask.  
\vspace{-2ex}
\subsection{Risk and performance metrics}
Each of the strategies is evaluated by a number of risk/performance metrics, especially including the following:
\begin{itemize}
    \item \textbf{Profit and Loss (PnL) distribution} statistics
    \item \textbf{Mean episode return}
    \item \textbf{Mean Absolute Position (MAP)}, approximated as:
        \begin{equation}
        \operatorname{MAP} = \frac{1}{N} \sum_{k=1}^{N} |I_{k\Delta t}|,
        \end{equation}
    where $N$ is the number of time-steps in an episode. 
    \item \textbf{Sharpe ratio}, which, in the context of high-frequency trading and MM, is commonly defined \cite{coates} as:
         \begin{equation}
        SR = \frac{\mu_{W_T}}{\sigma_{W_T}},
        \end{equation}
    where $\mu_{W_T}$ ($\sigma_{W_T}$) denotes the mean (standard deviation) of the terminal wealth (PnL).
    \item \textbf{(Mean PnL)/MAP} - the ratio of the mean terminal wealth (PnL) to the MAP. This is a non-rolling variant of the risk metric introduced in \cite{gasperov}.
    
\end{itemize}

\begin{table}
\centering
\scriptsize
\caption{Distributional statistics --- DRL controller vs benchmarks}
\vspace{-0.1cm}
\begin{tabular}[t]{lccccc}
\hline
Metric & DRL & SYM & LIN\\
\hline
Mean episode return & 13.3378 & 8.3913 & 7.6772 \\
Mean PnL & 13.7567 & 9.8686 & 8.2105 \\
Std PnL & 8.4952 & 8.2443 & 5.7884 \\
Kurtosis PnL & 4.2083 & 6.0552 & 10.5585 \\
Skew PnL & 1.5870 & 1.6375 & 2.3863 \\
Jarque Bera PnL & 578.8301 & 987.3005 & 2797.0786 \\
Jarque Bera PnL p-value (5\%) & 0 & 0 & 0\\
10th percentile PnL & 4.9445 & 1.6775 & 2.7400\\
20th percentile PnL & 7.0480 & 3.6220 & 3.9800\\
80th percentile PnL & 19.0340 & 15.0000 & 11.4360\\
90th percentile PnL & 24.4310 & 19.6150 & 14.7170\\
Sharpe Ratio & 1.6193 & 1.1970 & 1.4184  \\
Abs. mean terminal inv. & 0.454 & 1.46 & 0.56 \\
Mean terminal inv. & -0.122 & -0.028 & -0.028 \\
Std terminal inv. & 0.7477 & 1.7650 & 0.8070 \\
Kurtosis inv. & 1.6838 & -1.0332 & 0.0277  \\
Skew inv. & 0.3742 & 0.0032 & 0.1879  \\
Jarque Bera inv. & 70.7342 & 22.2388 & 2.9569\\
Jarque Bera inv. p-value (5\%) & 0 & 0 & 0.2280 \\
10th percentile inv. & -1 & -2 & -1 \\
20th percentile inv. & -1 & -2 & -1 \\
80th percentile inv. & 0 & 2 & 1 \\
90th percentile inv. & 1 & 2 & 1 \\
Mean Absolute Position (MAP) & 0.4232 & 1.4659 &
0.5478 \\
(Mean PnL)/MAP & 32.5064 & 6.7321 & 14.9881 \\

\hline
\end{tabular}
\label{table:table1}
\vspace{-2ex}
\end{table}%

\begin{figure}
\includegraphics[width=0.5\textwidth]{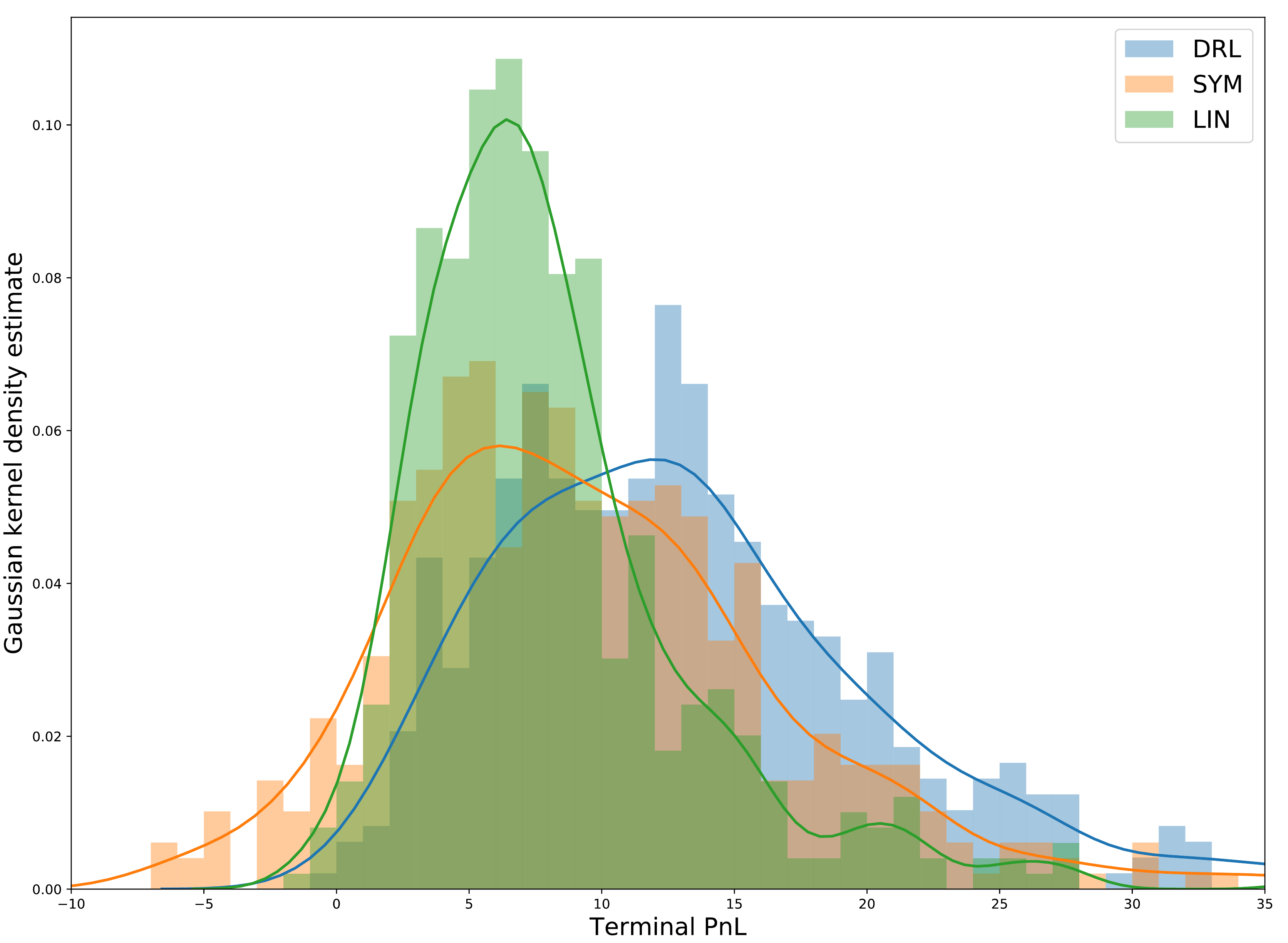}
\includegraphics[width=0.5\textwidth]{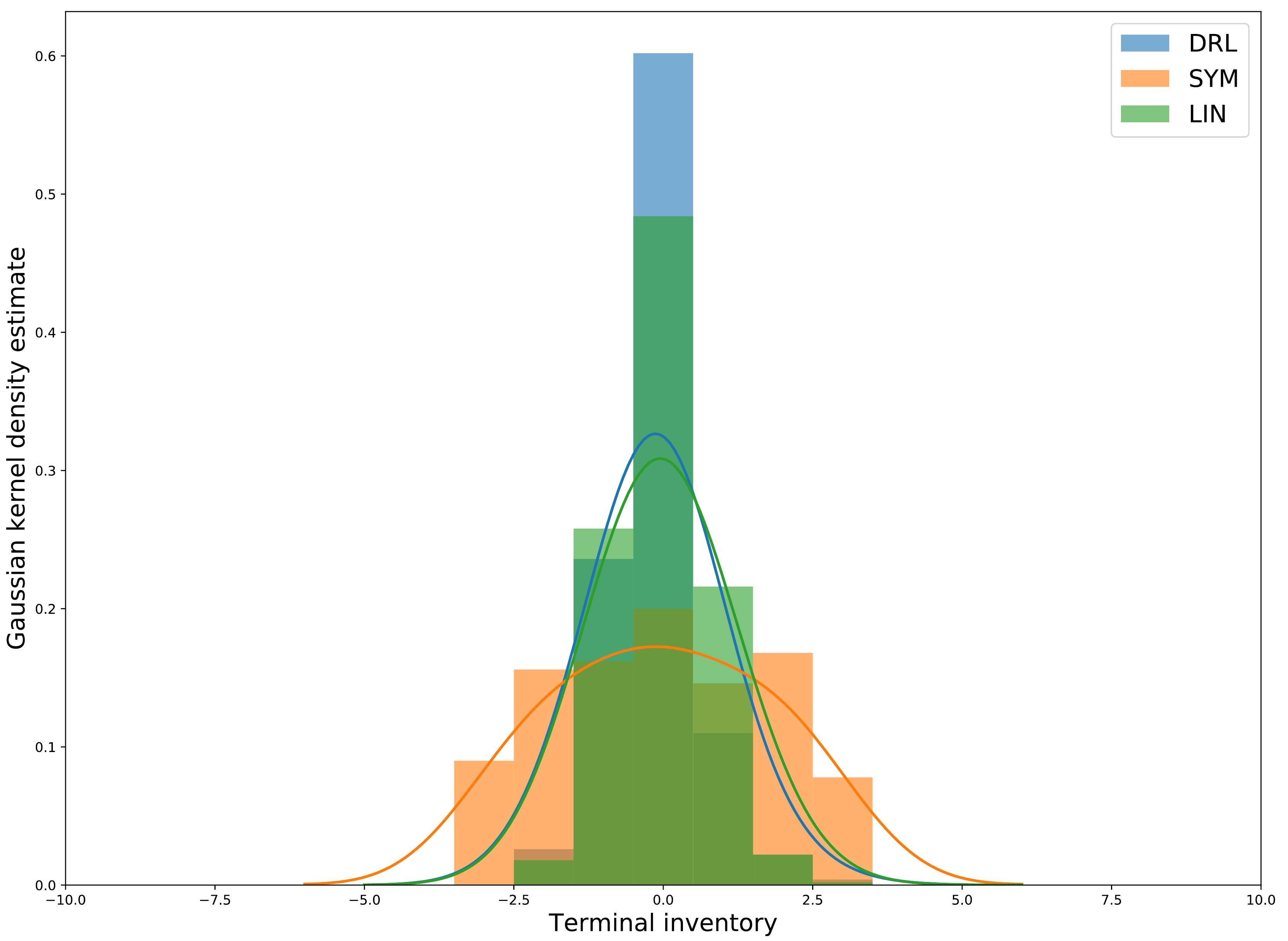}
\caption{PnL and terminal inventory distributions --- DRL versus benchmarks}
\label{fig:figure2}
\vspace{-4ex}
\end{figure}
\vspace{-3ex}
\subsection{Experimental results}
\begin{figure}[ht]
\vspace{0.5ex}
\includegraphics[width=0.5\textwidth]{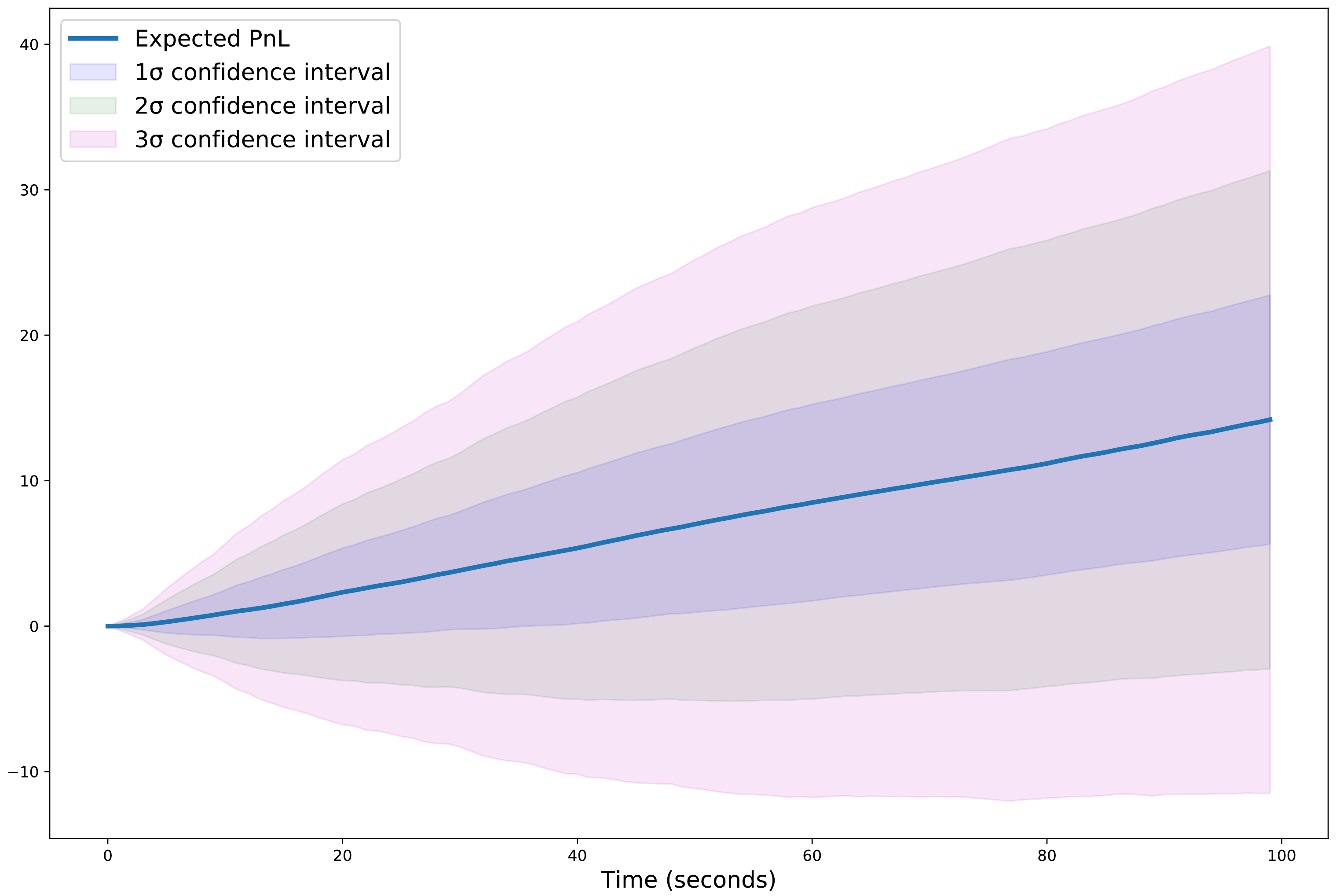}
\caption{DRL strategy --- mean PnL in time}
\label{fig:figure3}
\vspace{-4ex}
\end{figure}

The detailed results (distributional statistics for the PnL
and terminal inventory distribution as well as additional risk
metrics for all of the strategies) are provided in Table \ref{table:table1}. The strategies' terminal PnL and inventory distributions are shown in Fig. \ref{fig:figure2} together with Gaussian kernel density estimates. The results clearly indicate that the DRL strategy outperforms both of the benchmarks with respect to the vast majority of the considered metrics. The DRL strategy performance results, when juxtaposed with the SYM and LIN strategy, show a considerably higher mean PnL value as well as a more favorable Sharpe ratio. The percentiles of the PnL distribution (also interpreted as VaR) indicate its dominant performance as well. As expected, its MAP is fairly low, owing to the effect of inventory penalization, implying that the strategy succeeds at MM without exposing the trader to high inventory risks. Also note the significantly lower kurtosis value for its PnL distribution, indicating thinner tails, which is a desirable property for risk management purposes, as well as a somewhat smaller PnL distribution skewness. We also observe that the Jarque-Bera test results indicate inconsistency with the null hypothesis (normality of the PnL distribution) for all of the strategies. Interestingly, comparing the SYM and the LIN strategy, the latter exhibits lower risk at the expense of lower expected PnL, similarly to the results from the AS study \cite{as}. The terminal inventory distribution for all of the strategies is centered around zero, demonstrating telltale signs of MM behavior (position oscillating around zero inventory). We observe that the terminal inventory distributions corresponding to the DRL and the LIN strategy seem quite similar, indicating comparable risk preferences. On the contrary, notice the size of the absolute mean terminal inventory distribution corresponding to the SYM strategy, clearly indicating its naive disregard for the inventory risk. Fig. \ref{fig:figure3} depicts the mean performance of the DRL strategy in time together with the associated confidence intervals, with the mean PnL growing (seemingly) linearly in time.
\subsection{Sensitivity analysis}

Ideally, we would like to obtain MM controls robust to changes in the underlying order arrival intensity rates. Relatedly, we conduct sensitivity analysis of the DRL controller by changing the background intensity rates for all order types --– specifically by adding normal (Gaussian) noise. 
More precisely, three different noise sizes were considered --- Gaussian noise based with mean $0$ and variance $0.1$ (DRL-N-$0.1$), variance $0.2$ (DRL-N-$0.2$) and variance $0.3$ (DRL-N-0.3). The results are shown in Table \ref{table:table101}. Evidently, large perturbations increase the standard deviation of the terminal PnL distribution, while the effect on the mean is more complicated, with the largest noise size resulting in the largest mean PnL (and also MAP). Quite expectedly, the Sharpe ratio clearly decreases (while the MAP increases) as the size of the perturbation grows. 

\begin{table}[ht!]
\scriptsize
\vspace{-2ex}
\centering
\caption{Distributional statistics --- RL noise levels}
\begin{tabular}[t]{lcccc}
\hline
Metric & DRL-N-$0.1$ & DRL-N-$0.2$ & DRL-N-$0.3$ \\
\hline
Mean PnL & 13.2050 & 13.2553 & 15.5077 \\
Std PnL & 9.5922 & 13.0264 & 17.9523 \\
Kurtosis PnL & 4.5894 & 15.2773 & 12.0886 \\
Skew PnL & 1.6481 & 3.0457 & 2.9768 \\
Jarque Bera  PnL & 665.14 & 5635.43 & 3782.90 \\
Jarque Bera PnL p-value & 0 & 0 & 0 \\
10th percentile PnL & 3.7090 & 2.3745 & 1.9472\\
20th percentile PnL & 5.4676 & 4.4940 & 3.6387\\
80th percentile PnL & 19.0860 & 19.7860 & 22.2980\\
90th percentile PnL & 24.7940 & 26.3665 & 34.6055\\
Sharpe Ratio & 1.3766 & 1.0175 & 0.8638  \\
Mean Absolute Position & 0.4443 & 0.5038 & 0.5486 \\
\hline
\vspace{-0.5cm}
\end{tabular}
\label{table:table101}
\end{table}%

\begin{figure}
\vspace{-0.5ex}
\includegraphics[width=0.45\textwidth]{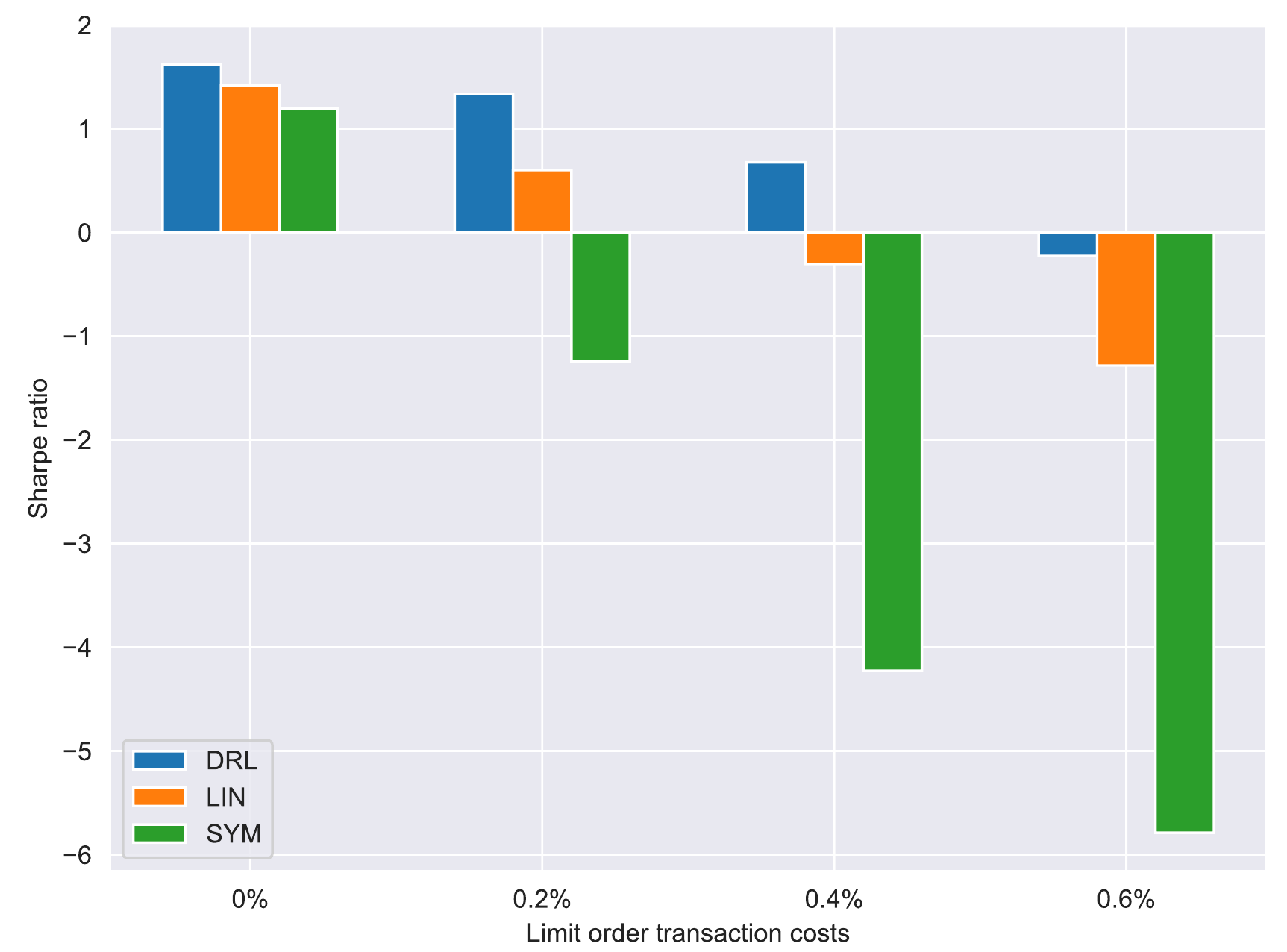}
\caption{Sharpe Ratio under variable limit order transaction fees}
\label{fig:figure5}
\vspace{-4ex}
\end{figure}

Furthermore, we investigate the performance of our strategy and the benchmark strategies under varying transaction costs. Fig. \ref{fig:figure5} shows the Sharpe ratio for the three strategies and variable limit order transaction fee rates. It is clear from the figure that the DRL strategy is capable of tolerating the inclusion of fees for limit order transactions, up to the fee of around $0.6 \%$. On the contrary, the LIN strategy generates negative profits already with fees set to $0.4 \%$, while the SYM strategy demonstrates extremely low tolerance to such transaction fees, even for the relatively low rate of $0.2 \%$. The considered rates lie in a realistic range, and therefore this validates the performance of our DRL method under real-life conditions. Finally, we retrain the DRL controller under varying limit order transaction fees and investigate the corresponding mean number of transactions. The mean number of transactions equals $24.54$, $15.0$, $11.28$ and $7.85$ under limit order transaction costs of $0 \%$, $0.2 \%$, $0.4 \%$, and $0.6 \%$, respectively. The results clearly indicate that higher limit order transaction costs, quite expectedly, lead to fewer realized transactions (trades), as the DRL controller becomes increasingly discriminating about when and if to post limit orders in the presence of high transaction costs.

\vspace{-1ex}
\section{CONCLUSION}
A DRL-based approach was used to obtain market-making strategies with superior performance, as compared to heuristic benchmarks. The approach yields promising results when realistic LOB simulators based on multivariate Hawkes processes are employed. Special focus was placed on the statistical analysis of the resulting PnL and terminal inventory distributions, and sensitivity analysis, both to changes in the underlying order intensity rates and the limit order fees. We conclude that DRL presents a viable approach for obtaining competitive MM controllers. Further research might consider more advanced MM models which take into account the full limit order book, or are based on Hawkes processes with alternative (power-law or polynomial) kernels. Finally, robust adversarial RL \cite{pinto} might be used to further improve generalization and robustness under model uncertainty.
\vspace{-2ex}
\section*{Appendix}
\vspace{-1ex}
The following hyperparameter values were used for training: $\gamma = 1$; batch size, $512$; buffer size, $10^{5}$; learning rate, $0.0003$; learning starts, $100$; ent. coef., "auto"; target update interval, $1$; gradient steps, $1$; train freq., $1$; $\tau=0.005$. The following MM procedure parameter values were used: $\Delta t = 1$; $T = 100$; $\delta = 0.01$; $\phi = 0.01$; $c=3$; $\mu=0.01$; $Z_1=\frac{8}{30}$; $Z_2=0.25$; $Z_3 = 0.25$; $\beta = 0.08$; market maker fee, $0\%$; market taker fee, $0.2\%$. The Stable Baselines package was used for the implementation. The code is provided at: \url{github.com/BGasperov/drlformm}
\vspace{-2ex}
\section*{Acknowledgments}
 We would like to thank the three anonymous reviewers and the editor for their extremely valuable and helpful comments.


\vspace{-4ex}
\begin{thebibliography}{00}
\bibitem[1]{as}
Avellaneda, M. and Stoikov, S., 2008. High-frequency trading in a limit order book. \emph{Quantitative Finance, 8}(3), pp.217-224.
\bibitem[2]{fodra}
Fodra, P. and Pham, H., 2015. High frequency trading and asymptotics for small risk aversion in a Markov renewal model. \emph{SIAM Journal on Financial Mathematics, 6}(1), pp.656-684.
\bibitem[3]{gueant}
Guéant, O., Lehalle, C.A. and Fernandez-Tapia, J., 2013. Dealing with the inventory risk: a solution to the market making problem. \emph{Mathematics and financial economics, 7}(4), pp.477-507.
\bibitem[4]{cartea}
Cartea, A., Jaimungal, S. and Ricci, J., 2018. Algorithmic trading, stochastic control, and mutually exciting processes. \emph{SIAM Review, 60}(3), pp.673-703.
\bibitem[5]{silver}
Silver, D., Huang, A., Maddison, C.J., Guez, A., Sifre, L., Van Den Driessche, G., Schrittwieser, J., Antonoglou, I., Panneershelvam, V., Lanctot, M. and Dieleman, S., 2016. Mastering the game of Go with deep neural networks and tree search. \emph{Nature, 529}(7587), pp.484-489.
\bibitem[6]{ranzato}
Ranzato, M.A., Chopra, S., Auli, M. and Zaremba, W., 2015. Sequence level training with recurrent neural networks. \emph{arXiv preprint arXiv:1511.06732}.
\bibitem[7]{burak}
Demirel, B., Ramaswamy, A., Quevedo, D.E. and Karl, H., 2018. Deepcas: A deep reinforcement learning algorithm for control-aware scheduling. \emph{IEEE Control Systems Letters, 2}(4), pp.737-742.
\bibitem[8]{lim}
Lim, Y.S. and Gorse, D., 2018, April. Reinforcement learning for high-frequency market making. In \emph{ESANN 2018-Proceedings, European Symposium on Artificial Neural Networks, Computational Intelligence and Machine Learning} (pp. 521-526). ESANN.
\bibitem[9]{cont}
Cont, R., Stoikov, S. and Talreja, R., 2010. A stochastic model for order book dynamics. \emph{Operations research, 58(3)}, pp.549-563.
\bibitem[10]{spooner}
Spooner, T., Fearnley, J., Savani, R. and Koukorinis, A., 2018. Market making via reinforcement learning. \emph{arXiv preprint arXiv:1804.04216.}
\bibitem[11]{sadighian}
Sadighian, J., 2019. Deep reinforcement learning in cryptocurrency market making. \emph{arXiv preprint arXiv:1911.08647}.
\bibitem[12]{gasperov}
Gašperov, B. and Kostanjčar, Z., 2021. Market Making With Signals Through Deep Reinforcement Learning. \emph{IEEE Access, 9}, pp.61611-61622.
\bibitem[13]{baldacci}
Baldacci, B., Manziuk, I., Mastrolia, T. and Rosenbaum, M., 2019. Market making and incentives design in the presence of a dark pool: a deep reinforcement learning approach. \emph{arXiv preprint arXiv:1912.01129}.
\bibitem[14]{gueant2}
Guéant, O. and Manziuk, I., 2019. Deep reinforcement learning for market making in corporate bonds: beating the curse of dimensionality. \emph{Applied Mathematical Finance, 26}(5), pp.387-452.
\bibitem[15]{ganesh}
Ganesh, S., Vadori, N., Xu, M., Zheng, H., Reddy, P. and Veloso, M., 2019. Reinforcement learning for market making in a multi-agent dealer market. \emph{arXiv preprint arXiv:1911.05892.}
\bibitem[16]{law}
Law, B. and Viens, F., 2019. Market making under a weakly consistent limit order book model. \emph{High Frequency, 2}(3-4), pp.215-238.
\bibitem[17]{bacry}
Bacry, E., Mastromatteo, I. and Muzy, J.F., 2015. Hawkes processes in finance. \emph{Market Microstructure and Liquidity, 1}(01), p.1550005.
\bibitem[18]{luabergel1}
Lu, X. and Abergel, F., 2018. High-dimensional Hawkes processes for limit order books: modelling, empirical analysis and numerical calibration. \emph{Quantitative Finance, 18}(2), pp.249-264.
\bibitem[19]{fonseca}
Da Fonseca, J. and Zaatour, R., 2014. Hawkes process: Fast calibration, application to trade clustering, and diffusive limit. \emph{Journal of Futures Markets, 34}(6), pp.548-579.
\bibitem[20]{deprado}
Lopez de Prado, M., 2019. Tactical investment algorithms. \emph{Available at SSRN 3459866.}
\bibitem[21]{embrechts}
Embrechts, P., Liniger, T. and Lin, L., 2011. Multivariate Hawkes processes: an application to financial data. \emph{Journal of Applied Probability, 48}(A), pp.367-378.
\bibitem[22]{law2}
Law, B. and Viens, F., 2016. Hawkes processes and their applications to high-frequency data modeling. \emph{Handbook of High-Frequency Trading and Modeling in Finance, 9}, p.183.
\bibitem[23]{biais}
Biais, B., Hillion, P. and Spatt, C., 1995. An empirical analysis of the limit order book and the order flow in the Paris Bourse. \emph{the Journal of Finance, 50}(5), pp.1655-1689.
\bibitem[24]{sornette}
Filimonov, V. and Sornette, D., 2015. Apparent criticality and calibration issues in the Hawkes self-excited point process model: application to high-frequency financial data. \emph{Quantitative Finance, 15}(8), pp.1293-1314.
\bibitem[25]{simon}
Simon, G., 2016. \emph{Hawkes Processes in Finance: A Review with Simulations} (Doctoral dissertation, University of Oregon).
\bibitem[26]{oakes}
Oakes, D., 1975. The Markovian self-exciting process. \emph{Journal of Applied Probability, 12}(1), pp.69-77.
\bibitem[27]{ogata}
Ogata, Y., 1981. On Lewis' simulation method for point processes. \emph{IEEE Transactions on Information Theory, 27}(1), pp.23-31.
\bibitem[28]{briola}
Briola, A., Turiel, J. and Aste, T., 2020. Deep Learning modeling of Limit Order Book: a comparative perspective. \emph{arXiv preprint arXiv:2007.07319.}
\bibitem[29]{gueantbook}
Guéant, O., 2016. \emph{The Financial Mathematics of Market Liquidity: From optimal execution to market making} (Vol. 33). CRC Press.
\bibitem[30]{coates}
Coates, J.M. and Page, L., 2009. A note on trader Sharpe Ratios. \emph{PloS one, 4}(11), p.e8036.
\bibitem[31]{pinto}
Pinto, L., Davidson, J., Sukthankar, R. and Gupta, A., 2017, July. Robust adversarial reinforcement learning. In \emph{International Conference on Machine Learning} (pp. 2817-2826). PMLR.

\end{thebibliography}
\end{document}